\documentclass[lettersize,journal]{IEEEtran}
\usepackage{amsmath,amsfonts}
\usepackage{algorithmic}
\usepackage{algorithm}
\usepackage{array}
\usepackage[caption=false,font=normalsize,labelfont=sf,textfont=sf]{subfig}
\usepackage{textcomp}
\usepackage{stfloats}
\usepackage{url}
\usepackage{verbatim}
\usepackage[none]{hyphenat} 
\usepackage{graphicx}
\usepackage{cite}
\begin{document}

\title{\huge{Quasi-Closed-Form Driven Near-Field Flat-Top Beamfocusing with Concentric Circular Vertical Polarized Dipole Array For Large Intelligent Surface Applications}}

\author{{Jiawang Li,~\IEEEmembership{Student Member,~IEEE}}
\thanks{Manuscript received xxxx.xx.xx. (Corresponding author: \textit{Jiawang Li}).}
\thanks{Jiawang Li is with the Department of Electrical and 
Information Technology, Lund University, 22100 Lund, Sweden 
(e-mail: {jiawang.li@eit.lth.se}).}
\thanks{Color versions of one or more of the figures in this communication are available online at http://ieeexplore.ieee.org.}}

\markboth{JOURNAL OF LATEX CLASS FILES, VOL. 14, NO. 8, FEBRUARY 2025}%
{Shell \MakeLowercase{\textit{et al.}}: A Sample Article Using IEEEtran.cls for IEEE Journals}


\maketitle

\begin{abstract}
This letter presents a near-field flat-top beam synthesis method based on a semi-closed-form approach. First, the feasibility of achieving a flat-top beam in the near field is examined using a closed-form analysis. A circular concentric ring array structure is adopted, and it is observed that circular rings with different radii exhibit distinct gain characteristics along the focal region on the $z$-axis. Specifically, smaller radii lead to a monotonic increase in electric field strength near the focus, whereas larger radii result in a monotonic decrease. Based on this behavior, parameters such as the number of rings and the initial radius are determined through field superposition. Subsequently, an optimization algorithm is employed to fine-tune the excitation amplitudes of the individual rings in order to suppress sidelobes. The effectiveness of the proposed method is validated through full-wave electromagnetic simulations.
\end{abstract}


\section{Introduction}
\IEEEPARstart{T}{his} With the rapid development of wireless communications, antennas play an increasingly important role in fulfilling tough performance requirements. In many applications, specific radiation patterns in far field are needed, such as high-gain pattern from phased array antennas for 5G small base stations [1], monopole-like pattern for indoor WLAN [2], flat-top pattern for microwave power transmission [3] and tilted beam for smart transportation [4]. 
In recent years, the emergence of large-scale multiple-input multiple-output (MIMO) systems has led to significant advances in wireless communications. A particularly promising architecture within this domain is the large intelligent surface (LIS) [5], which extends the traditional concept of antenna arrays to massive, two-dimensional, surface-based deployments. With a substantial increase in the number of antenna elements, LIS-based MIMO systems are capable of fine-grained spatial control, opening new possibilities in energy focusing, spatial multiplexing, and interference management.

As the physical dimensions of LIS grow, the assumption of far-field signal propagation becomes increasingly invalid, especially when the transmission distance falls within the radiating near-field region. This shift has sparked considerable research interest in near-field communications, where wavefronts must be treated as spherical rather than planar [6], and beamforming strategies must incorporate both angle and distance dependence [7]–[13].
\begin{figure}[!t]
\centering
\includegraphics[width=2.6in]{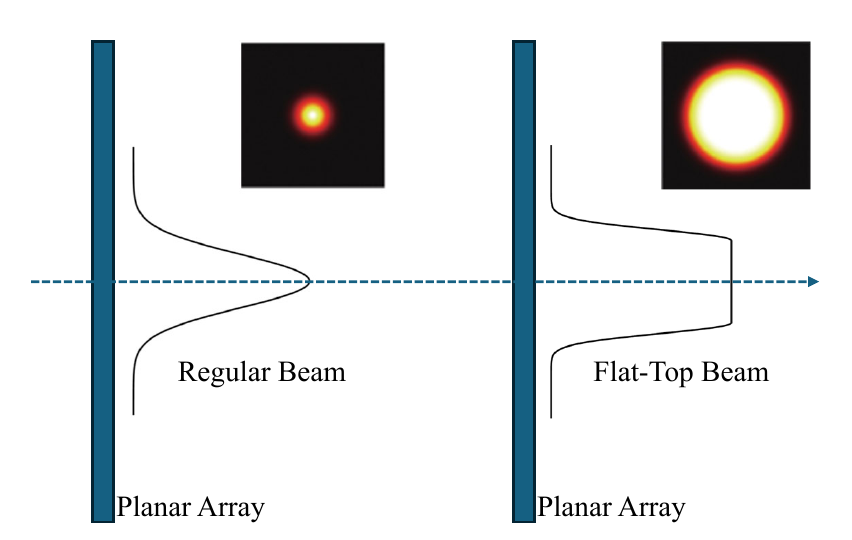}
\caption{Near-field regular beam and flat-top beam.}
\label{fig1}
\end{figure}
Among the various innovations in near-field beamforming, the design of flat-top beams has attracted attention due to its potential to deliver uniform power distribution over a specified spatial region. Unlike conventional beams that focus energy at a single point, flat-top beams aim to maintain a relatively constant field intensity across a targeted area.  
As shown in Fig. 1, two distinct near-field beam patterns are presented. The conventional beam provides peak gain at the focal point but exhibits rapid attenuation with longitudinal displacement. In contrast, the flat-top beam maintains stable gain over a defined range along the propagation axis. This feature is particularly beneficial in applications such as wireless power transfer, where it ensures consistent charging for multiple devices at the same distance and improves efficiency for receivers with antenna arrays. It also offers reliable coverage for slightly moving targets by reducing sensitivity to positional shifts. 

This paper explores the theoretical formulation and practical implementation of near-field flat-top beamforming for LIS-aided MIMO systems. A novel beam synthesis framework is proposed to generate spatially uniform energy distributions within a designated near-field zone. The performance of the proposed method is evaluated through full-wave simulations, demonstrating its effectiveness in delivering uniform wireless power transmission in near-field scenarios.

\section{ANTENNA STRUCTURE AND DESIGN PROCEDURE}
\subsection{The Problem Formulation}
\begin{figure}[!t]
\centering
\includegraphics[width=2.6in]{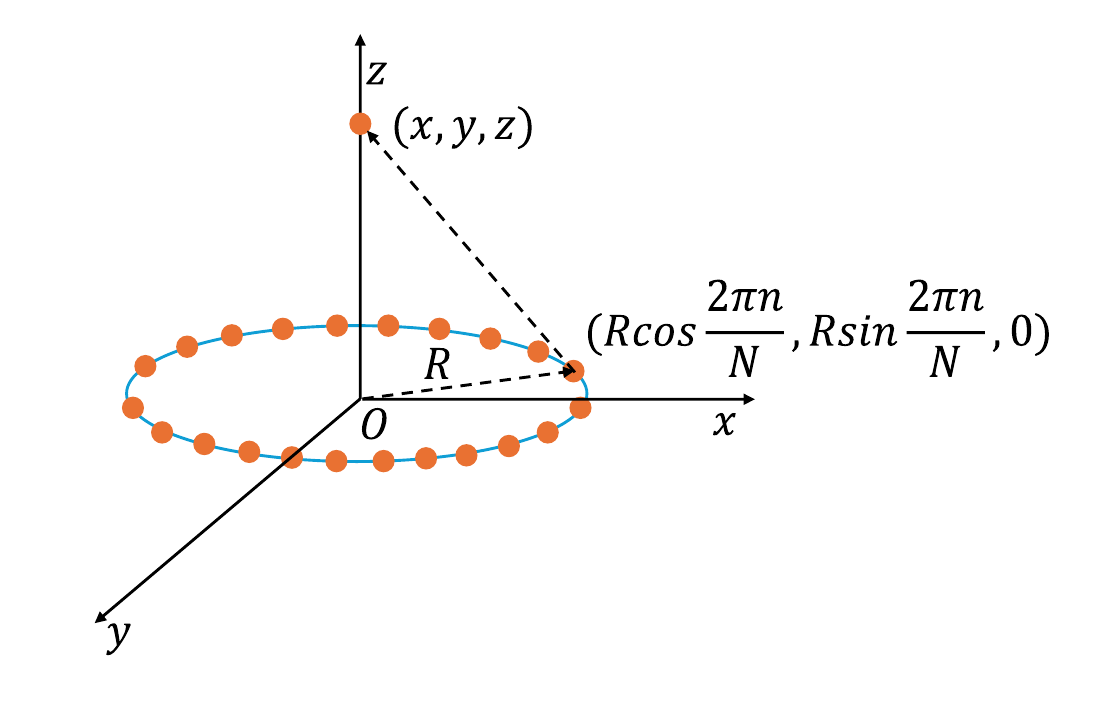}
\caption{Configuration of the proposed circular array.}
\label{fig1}
\end{figure}
As shown in Fig. 2, for a circular array, it is assumed that the spacing between adjacent elements is 
half a wavelength. This array is used as a transmitting array. For simplicity, the number of elements \( N \) 
is assumed to be even. The coordinates of the array elements in 3D space can be expressed as 
\( (R\cos(2\pi n/N), R\sin(2\pi n/N), 0) \). 
The coordinates of the receiving antenna are \( l = (x, y, z) \). 
All antennas, including the receiving antenna, are dipoles and share the same polarization, 
with the polarization direction parallel to the z-axis. Therefore, the polarization vector 
can be expressed as \( \hat{s} = (0, 0, 1) \). 
Taking the center of the circular array as the origin of the coordinate system, and neglecting the effects 
of spatial scattering and mutual coupling between antenna elements, the vertical electric field along the $\it{z}$-axis can be expressed as:
\begin{equation}
\bar{E}(r) = \frac{j\eta I_0 l k}{4 \pi  r} 
\left( \frac{\vec{r}}{|\vec{r}|} \times \left( \hat{s} \times \frac{\vec{r}}{|\vec{r}|} \right) \right) e^{-jkr},
\end{equation}
here, $k$ represents the wavenumber, defined as $k = \frac{2\pi}{\lambda}$, where $\lambda$ is the wavelength. 
$L$ denotes the effective length of the dipole, and $I_0$ is the current of the dipole. 
$\varepsilon_0$ represents the permittivity of vacuum, and $r$ is the distance from the focal point to the center of the dipole element, which can be expressed as:
\begin{equation}
|\vec{r}| = |\vec{l} - \vec{R}|.
\end{equation}
Then we can according to the following:
\begin{equation}
\left( \frac{\vec{r}}{|\vec{r}|} \times \hat{s} \right) \times \frac{\vec{r}}{|\vec{r}|} 
= \hat{s} - \left( \frac{\vec{r}}{|\vec{r}|} \cdot \hat{s} \right) \frac{\vec{r}}{|\vec{r}|},
\end{equation}
Thus, (1) can be simplified as follow:
\begin{equation}
E_x = \frac{j\eta I_0 L k}{4 \pi } 
\sum_{n=1}^{N} z|x-R\cos \frac{2\pi n}{N}| Q(x,y,z),
\end{equation}
\begin{equation}
E_y = \frac{j\eta I_0 L k}{4 \pi } 
\sum_{n=1}^{N} z|y-R\sin \frac{2\pi n}{N}| Q(x,y,z),
\end{equation}
\begin{equation}
E_z = \frac{j\eta I_0 L k}{4 \pi } 
\sum_{n=1}^{N} P(x,y) Q(x,y,z),
\end{equation}
$\left| \cdot \right|$ for 
$E_x$ and $E_y$ can ensure that the electric fields at the focal point are coherently added in phase. Where $Q(x,y,z)$ is expressed as:
\begin{equation}
P(x,y) = \left( x - R \cos \frac{2\pi n}{N} \right)^2 + \left( y - R \sin \frac{2\pi n}{N} \right)^2,
\end{equation}
\begin{equation}
Q(x,y,z) = \frac{e^{-jk \sqrt{P(x,y) + z^2}}}
{\sqrt{P(x,y) + z^2}^3},
\end{equation}
By normalizing the amplitude component of the electric field \( E_z \),
assuming \( \frac{j\eta I_0 L k}{4\pi} = 1 \),
and setting the focal point at \( (0, 0, z_0) \),
the symmetry of the system allows the intensity distribution at the focal point 
along the $x$-axis or $y$-axis to be expressed as:
\begin{equation}
E_z = \sum_{n=1}^{N} 
\frac{\left( \delta^2 - 2\delta R \cos \frac{2\pi n}{N} + R^2 \right) 
\phi(\delta)}
{\sqrt{\delta^2 - 2\delta R \cos \frac{2\pi n}{N} + R^2 + z_0^2}^3}.
\end{equation}
where $\phi(\delta) = e^{-jk \left( \sqrt{R^2 + z_0^2} - \sqrt{\delta^2 - 2\delta R \cos \frac{2\pi n}{N} + R^2 + z_0^2} \right)}$. Using the Taylor series expansion to simplify the exponential term and considering that the circular array has a sufficiently large number of elements such that the discrete elements can be approximated as a continuous distribution, equation (5) can be rewritten as:
\begin{equation}
E_z \approx \int_{0}^{2\pi} 
\frac{\left( \delta^2 - 2\delta R \cos \theta + R^2 \right) 
e^{-jk \left( \frac{\delta R \cos \theta}{\sqrt{R^2 + z_0^2}} \right)}}
{\sqrt{\delta^2 - 2\delta R \cos\theta + R^2 + z_0^2}^3}
\, d\theta.
\end{equation}
Consequently, a closed-form expression can be derived in terms of the focal position \( z_0 \), the radius of the circular array \( R \), and the wavenumber \( k \), which is related to the frequency:
\begin{equation}
\begin{split}
E_z = \left| 
2\pi \frac{(\delta^2 + R^2)}{\sqrt{R^2 + z_0^2}^3} 
J_0 \left( \frac{\delta k R}{\sqrt{R^2 + z_0^2}} \right) \right. \\
\left. - \frac{4i\pi \delta R}{\sqrt{R^2 + z_0^2}^3} 
J_1 \left( \frac{\delta k R}{\sqrt{R^2 + z_0^2}} \right) 
\right|.
\end{split}
\end{equation}
Consider two cases. In the first case, when R is relatively small, equation (7) can be approximated as:
\begin{equation}
E_z \approx \left| 
2\pi \frac{(\delta^2 + R^2)}{ \sqrt{R^2 + z_0^2}^3} 
\right|.
\end{equation}
\begin{figure}[!t]
\centering
\includegraphics[width=3in]{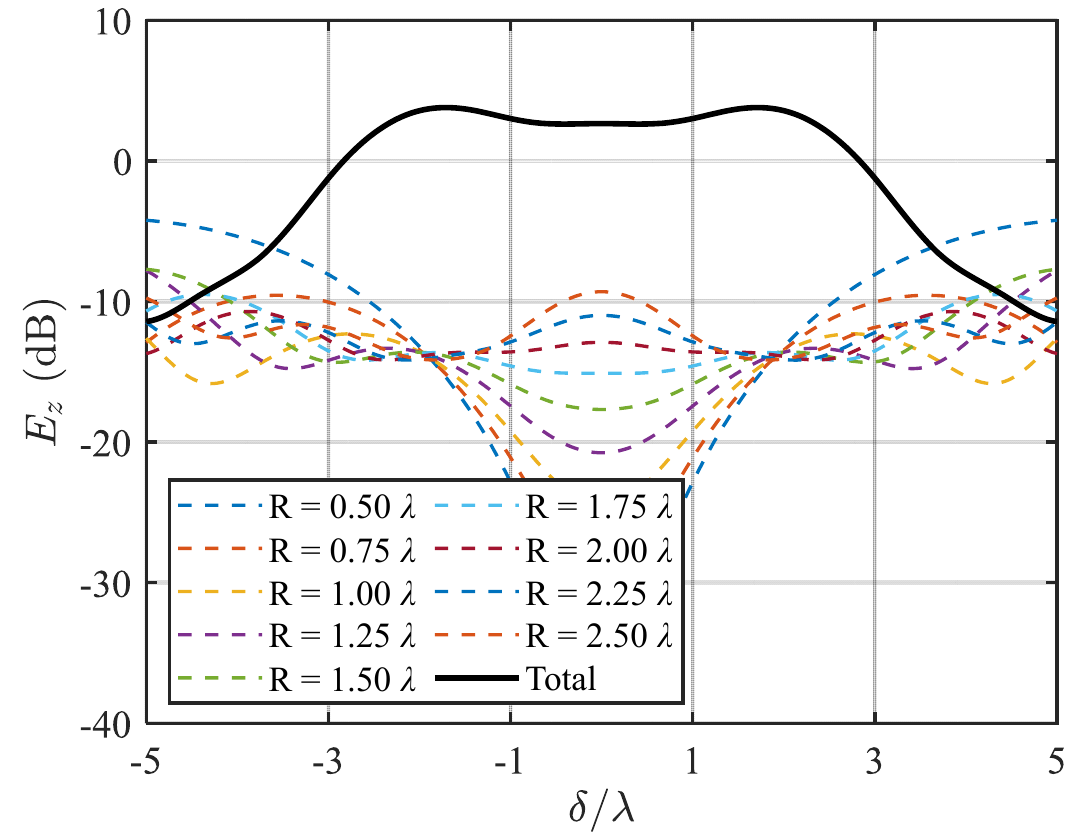}
\caption{$E_z$ changes under the condition of $z_0 = 10 \lambda$.
}
\label{fig1}
\end{figure}
At this point, it is evident that the function is an increasing function of \( \delta \) when \( R \) is small 
and \( \delta \) is also small. This is because one term, related to the Bessel function, approaches 1, 
while the other term exhibits a clear increasing trend. 
Next, consider the case when \( R \) is not small. In this case, equation
\begin{algorithm}[H]
\caption{Flat-top Field Synthesis via Circular Ring Array Search}
\label{alg:flat_top_synthesis}
\begin{algorithmic}[1]
\REQUIRE Wavelength $\lambda$, focus distance $z_0$, minimum radius $R_{\min}$, max rings $R_{\max}$, fluctuation tolerance $\rho$, scan range $\delta \in [-\delta_{\max}, \delta_{\max}]$
\STATE \textbf{Initialize:} Define candidate element spacings $d \in [\lambda/4, \lambda/2]$
\FORALL{spacing $d$ in candidate list}
    \FOR{$R_n = 3$ to $R_{\max}$}
        \STATE Compute ring radii $R = R_{\min} + d \cdot (0:R_n-1)$
        \STATE Set $N_1 \leftarrow \left\lfloor \frac{2\pi R_1}{d} \right\rfloor$, and $N_m \leftarrow N_1$ for all $m$
        \STATE Initialize total field $E_{\text{total}} \leftarrow 0$
        \FOR{$m = 1$ to $R_n$}
            \STATE Compute angles $\theta_n = 2\pi n / N_m$
            \FORALL{$\delta$ in scan range}
                \STATE Compute each element contribution to field $E_{m}(\delta)$ using phase and distance terms
            \ENDFOR
            \STATE $E_{\text{total}} \leftarrow E_{\text{total}} + E_m$
        \ENDFOR
        \STATE Compute dB pattern: $E_{\text{dB}} \leftarrow 20\log_{10}(|E_{\text{total}}|)$
        \STATE Compute fluctuation: $\Delta E \leftarrow \max(E_{\text{dB}}) - \min(E_{\text{dB}})$
        \IF{$\Delta E \leq \rho$}
            \STATE \textbf{Output:} optimal spacing $d^*$ and ring count $R_n^*$
            \STATE \textbf{break both loops}
        \ENDIF
    \ENDFOR
\ENDFOR
\IF{no configuration found}
    \STATE \textbf{Return:} "No feasible design under constraints"
\ENDIF
\end{algorithmic}
\end{algorithm}
(9) can be approximated as:
\begin{equation}
E_z = \left| 
2\pi \frac{ R^2}{\sqrt{R^2 + z_0^2}^3} 
J_0 \left( \frac{\delta k R}{\sqrt{R^2 + z_0^2}} \right)
\right|.
\end{equation}
The term associated with the Bessel function exhibits a decreasing trend before the first zero of the Bessel function. 
Although the other term shows an increasing trend, its growth is relatively slow. 
As a result, the Bessel function dominates the overall trend, leading to a decreasing trend for small values of \( \delta \). 

The differing electric field trends observed for different radii provide an opportunity for optimization. 
Specifically, circular arrays with varying radii can be combined to leverage the increasing trend for smaller radii 
and the decreasing trend for larger radii. This superposition can generate a flat-top beam. 
Moreover, since the Bessel function decreases more rapidly, this characteristic helps suppress energy spread outside 
the flat-top beam to some extent. 

To validate this concept, a simulation is conducted with \( z_0 = 10\lambda \), (all simulations in this work were conducted using a wavelength of 0.2m, corresponding to a center frequency of 1.5GHz), the focal point located at \( (0, 0, z_0) \), 
and 9 concentric circular arrays with radii ranging from 0.1 m to 0.9 m. Each circular ring is equipped with 12 vertically oriented dipole elements, rather than being uniformly spaced at half-wavelength intervals. This configuration ensures that the normalized gain defined in (7) remains consistent across rings, as will be discussed in detail later.
The electric field gain curves for each circular array, as well as the resulting superimposed curve, are shown in Fig. 3. 
A flat-top beam can be generated along the $\it{x}$- or $\it{y}$-axis. The formation of the flat-top beam is fundamentally attributed to the differing gain trends exhibited by concentric rings of varying radii within the focal region. Specifically, rings with smaller radii produce a slight central gain depression, while those with larger radii generate a slight central gain elevation. This compensatory characteristic leads to the beam flattening effect.
 The flat-top beam spans approximately from $-1.5\lambda$ to $1.5\lambda$, with a gain ripple of less than 1\,dB within this region.
\begin{figure}[!t]
\centering
\includegraphics[width=3in]{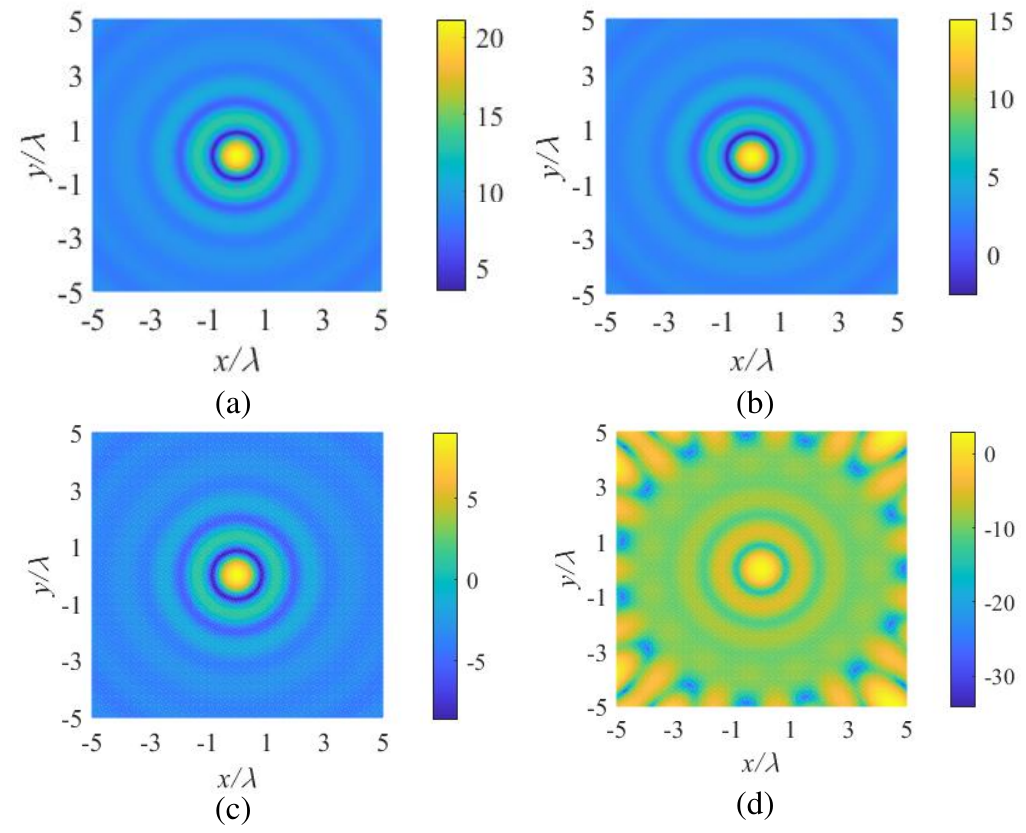}
\caption{Beam focusing results under the conjugate matching method with a radius of $5\lambda$ and a focal point located at $(0, 0, 10\lambda)$, for different element spacings. (a) $\lambda/4$. (b) $\lambda/2$. (c) $\lambda$. (d) $2\lambda$.
}
\label{fig1}
\end{figure}
\begin{figure}[!t]
\centering
\includegraphics[width=3.3in]{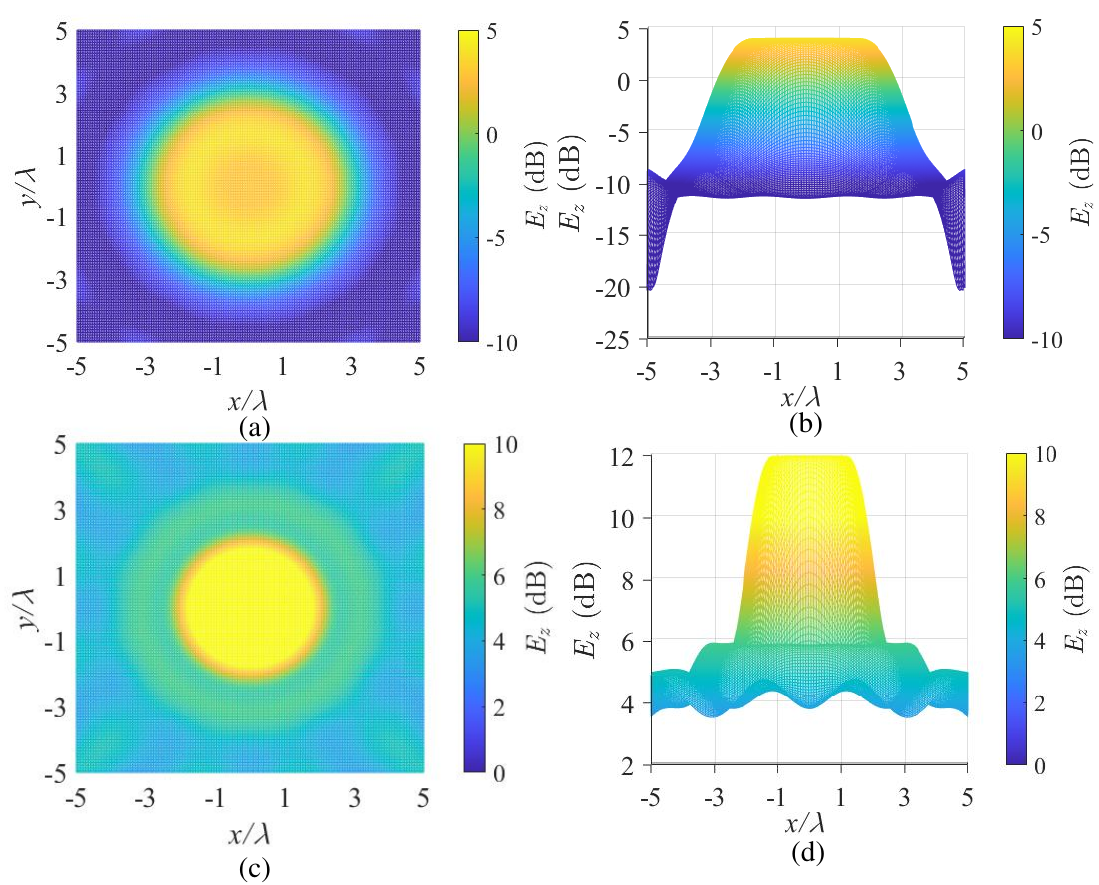}
\caption{Flat-top beam patterns synthesized by two algorithms. (a) 9 rings with a focal distance of $10\lambda$; (b) 6 rings with a focal distance of $5\lambda$. Each ring has 12 elements.
}
\label{fig1}
\end{figure}
It is important to note that each circular array must have a sufficient number of elements to ensure that 
the resulting beam includes the term associated with the Bessel function. 
As shown in Fig. 4, the beam distribution under different element counts is analyzed. 
The simulation setup is based on a circular array with a radius of 1 m and a focal point at \( (0, 0, 2) \). 
When the element spacing is less than \( 2\lambda \), the transverse gain variation at the focal point remains largely consistent. 
However, when the number of elements is further reduced, resulting in an element spacing of \( 3\lambda \), 
the beam energy at the sides of the array increases, leading to a significant relative reduction in the main beam gain. 
Therefore, it is crucial to select an appropriate number of elements to minimize the formation of large grating lobes 
or sidelobes and maintain optimal beam performance.

Algorithm 1 outlines the complete procedure. By initializing the wavelength, focal distance, minimum ring radius, maximum number of rings, and desired in-region flatness, the algorithm efficiently identifies a solution through a flatness criterion. This approach reduces the conventional per-element amplitude and phase search to a two-variable optimization over element spacing and ring count. It ensures uniform excitation and allows rapid phase allocation via conjugate matching based on the focal point location.

\section{Algorithms and FEKO Verification}

In this section, two examples are implemented based on the proposed algorithm. To verify its correctness, one of the cases is simulated using full-wave electromagnetic analysis in FEKO~2022. All algorithmic computations were performed on a workstation with an Intel i7-13700K CPU and 64\,GB RAM, and the execution time was recorded for performance evaluation.
As shown in Fig. 5, flat-top beam results under two different focal conditions are presented. In the first case, the focal point is located at a distance of $10\lambda$, while in the second it is positioned at $5\lambda$. Both configurations employ identical element spacing and inter-ring spacing of $\lambda/4$. However, the array in the first case consists of 9 concentric rings, whereas the second contains only 6. The corresponding execution times for the algorithm are 0.040\,s and 0.022\,s, respectively.

In both cases, the in-region flatness was constrained to be less than 1\,dB. Although the resulting flat-top beams exhibit different flat-top region radii---approximately $\pm 2\lambda$ in the first case and $\pm 1.2\lambda$ in the second---the second beam demonstrates higher radiated power due to the shorter focal distance. The peak electric field within the flat-top region reaches 11.94\,dB, with a gain fluctuation of only 0.51\,dB.

Similarly, Fig.6 and Fig.7 present the results of the synthesized flat-top beam in terms of the $E_y$ component, while the corresponding FEKO simulation results are shown in Fig. 8. Each dipole has a length of 100~mm, corresponding exactly to a half-wavelength. Ideal conductors are used for excitation, without considering finite radius or feed gap. Notably, during phase assignment, opposite signs are applied to the dipoles located at positions with $y > 0$ and $y < 0$, respectively. This ensures constructive interference at the focal point rather than signal cancellation.
It can be observed that, due to mutual coupling effects among array elements in practical implementations, the beamwidth is stretched along the flat-top direction, whereas the beamwidth in the orthogonal direction remains largely unaffected. It is worth noting that the beam broadening leads to a reduction in flatness within the targeted region, decreasing to approximately 4~dB. This distortion, induced by mutual coupling, primarily arises from the fact that the closed-form expression assumes ideal radiation patterns for each dipole element. In practice, however, coupling effects degrade the ideality of individual element patterns, resulting in discrepancies between the analytical predictions and full-wave numerical simulations. Nonetheless, a consistent trend between the two approaches can still be observed, which provides a useful foundation for the rapid synthesis of flat-top beamforming schemes.

\begin{figure}[!t]
\centering
\includegraphics[width=3in]{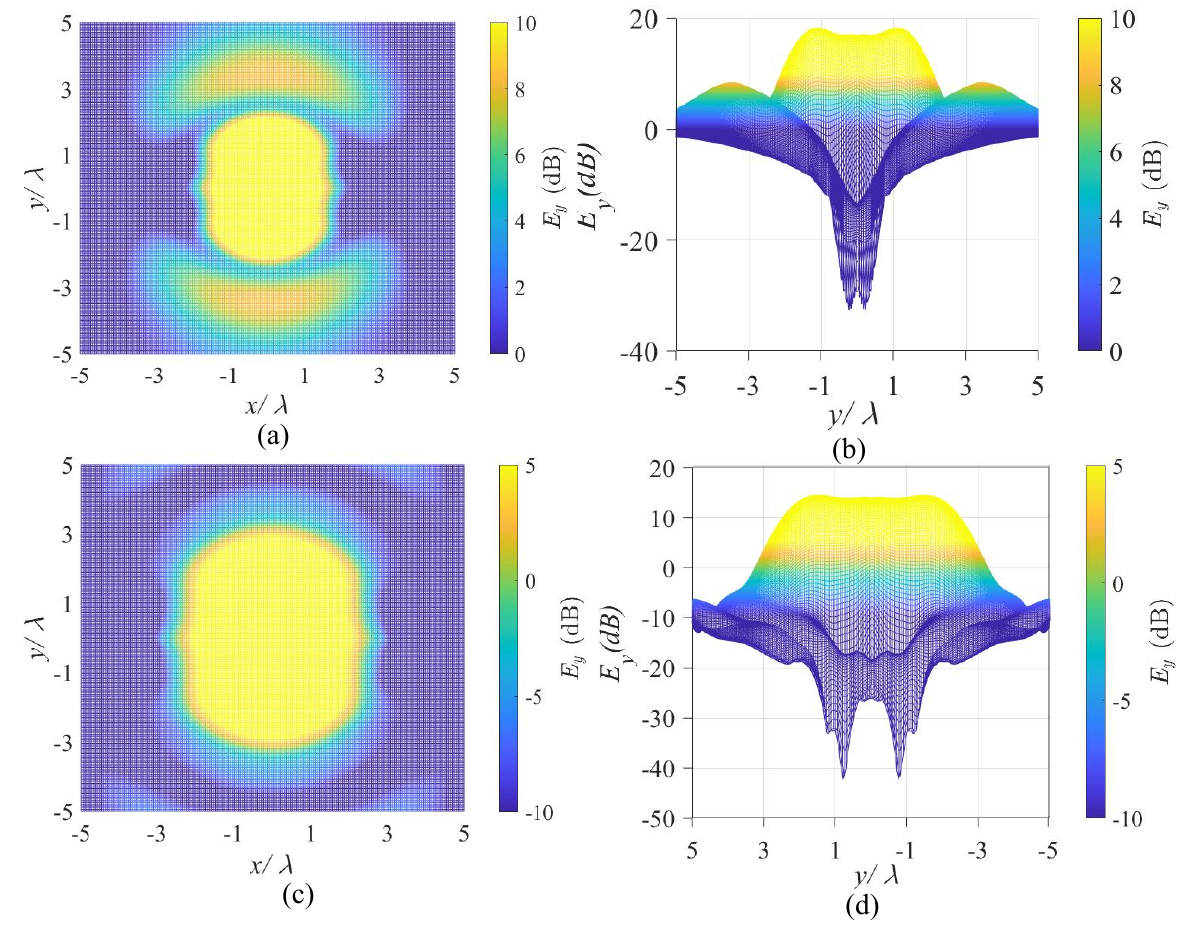}
\caption{Flat-top beam patterns synthesized by two algorithms. (a) 9 rings with a focal distance of $5\lambda$; (b) 5 rings with a focal distance of $10\lambda$. Each ring has 12 elements.
}
\label{fig1}
\end{figure}
\begin{figure}[!t]
\centering
\includegraphics[width=3in]{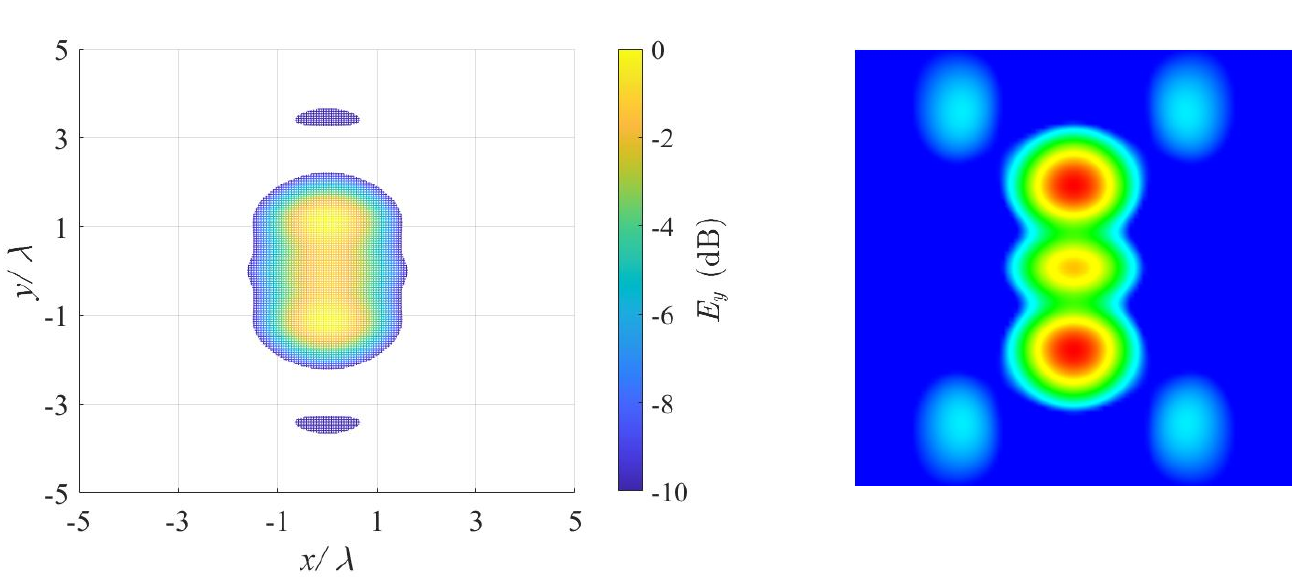}
\caption{Normalized simulated beams (-10 to 0 dB): (a) numerical simulation; (b) FEKO simulation.}
\label{fig1}
\end{figure}
\section{Conclusion}
This paper presents a near-field flat-top beam synthesis method based on a semi-closed-form approach. By analyzing the distinct field contributions of concentric circular rings with different radii, key design parameters such as the number of rings and the initial radius are determined through field superposition. The method reduces optimization complexity by converting per-element tuning into a low-dimensional search over global parameters. Uniform excitation is maintained, and phase assignment is efficiently handled via conjugate matching. Full-wave simulations validate the effectiveness of the proposed approach.

\vfill

\end{document}